# Quantum Tomography of the Photon-Plasmon Conversion Process in a Metal Hole Array

LEI TANG,[1,†] KAIMIN ZHENG,[1,†] JIALE GUO,[1] YI OUYANG,[1] YANG WU,[1] CHUANQING XIA,[1] LONG LI,[1] FANG LIU,[1] YONG ZHANG,[1,*] LIJIAN ZHANG,[1,*] AND MIN XIAO[1,2,*]

[1]*National Laboratory of Solid State Microstructures, College of Engineering and Applied Sciences, and School of Physics, Nanjing University, Nanjing 210093, China*
[2]*Department of Physics, University of Arkansas, Fayetteville, Arkansas 72701, USA*
[†]*These authors contribute equally to this work*
*\*zhangyong@nju.edu.cn,*
*\*lijian.zhang@nju.edu.cn,*
*\*mxiao@uark.edu*

**Abstract:** In the past decades, quantum plasmonics has become an active area due to its potential applications in on-chip plasmonic devices for quantum information processing. However, the fundamental physical process, i.e., how a quantum state of light evolves in the photon-plasmon conversion process, has not been clearly understood. Here, we report a complete characterization of the plasmon-assisted extraordinary optical transmission process through quantum process tomography. By inputting various coherent states to interact with the plasmonic structure and detecting the output states with a homodyne detector, we reconstruct the process tensor of the photon-plasmon conversion process. Both the amplitude and phase information of the process are extracted, which explains the evolution of the quantum-optical state after the coupling with plasmons. Our experimental demonstration constitutes a fundamental block for future on-chip applications of quantum plasmonic circuits.

## 1. Introduction

Plasmonics provides the capabilities to localize and manipulate electromagnetic excitations within sub-wavelength scales, and therefore has great potentials in the miniaturization and scalability of photonic devices[1-5]. In recent years, the newly emerging field of quantum plasmonics[6,7] has become an active research area, which is mostly motivated by its potential applications in integrated plasmonic circuits for quantum information processing. Quantum plasmons excited by various quantum sources including single photon[8-10], two-photon entanglement[11,12] and squeezing[13] have been experimentally demonstrated. The fundamental physical characteristics of quantum plasmons have been investigated in intense light-matter interactions[14-16], quantum confinement effects[17], survival of entanglement and squeezing[11,18-20], decoherence and loss[8,11,18], perfect absorption of entangled photons[21] etc. Quantum plasmonic devices such as detectors[22,23], interferometers[9,24,25], and controlled-NOT gate[26] have been developed. The applications have been further extended to quantum plasmonic sensing and quantum plasmonic networks beyond the classical limit[27,28]. However, large-scale usage of quantum plasmons in quantum information processing still faces many technical challenges to be solved, such as reducing loss and enhancing efficiency of plasmon-based quantum optical logic gates, controlling the phase of quantum interference devices and understanding at what scale the quantum theory has to be considered in a plasmonic device. More importantly, the fundamental physical problem, how an arbitrary quantum state of light evolves through the coupling with plasmons, has not been investigated systematically. Physical understanding of quantum plasmons can be acquired in two steps through the evolution of various degrees of freedom of the optical field during the photon-plasmon conversion process. The first one is about the degree of freedom of single photons, e.g. polarization, spatial, temporal and spectral, which have been studied extensively[11,18,29]. However, the quantum fluctuations of photon-number degree of freedom and, moreover, the quasi-probability distribution in the phase space mark the difference between classical and non-classical optical fields, and are crucial for various quantum information applications[27,30,31]. Therefore, a complete understanding of the evolution of quantum fluctuations and coherence inside the plasmonic structure is indispensable.

In this work we address this problem by applying the coherent-state quantum process tomography[32] (CSQPT) to a typical plasmon-assisted process, i.e. extraordinary optical transmission[33,34](EOT), in a metal-hole array. This transmission is largely attributed to the excitation of surface plasmon polaritons (SPPs) in the literature. Although subsequent theoretical and experimental work has claimed the presence of other surface waves[35] in this process, SPPs still play an important role in EOT process at visible and near-infrared frequencies. And it has been shown that such process can maintain certain quantum properties, for example, two-photons entanglement[11] and squeezing[13]. Here, we focus the character of far-fields in EOT process, by inputting a set of coherent states through a metal hole array and performing a tomographic reconstruction of the output states, we provide a complete characterization of the plasmon-assisted EOT process that spans the effect of the process in the photon-number Hilbert space. In particular, the reconstruction allows to predict the output state of any input states including non-classical ones.

## 2. Coherent state quantum process tomography

A quantum process $\varepsilon$ can be described by a positive, trace-preserving linear map that transfers the input states to the output states over Hilbert space H. Complete characterization of a quantum process therefore means to know the effect of the process on arbitrary quantum states. Assuming $\{\hat{\rho}_i\}$ is the complete set spanning the single-mode Hilbert space, any input state can be decomposed as

$$\hat{\rho}^{in} = \sum_i a_i \hat{\rho}_i. \tag{1}$$

Therefore complete characterization of a quantum process is equivalent to determine the output state $\varepsilon(\hat{\rho}_i)$ for each $\hat{\rho}_i$. Once this information is acquired, one can predict the output state $\hat{\rho}^{out} = \varepsilon(\hat{\rho}^{in})$ through

$$\hat{\rho}^{out} = \sum_i a_i \varepsilon(\hat{\rho}_i). \tag{2}$$

The challenge associated with this approach is the construction of the appropriate complete set. For the optical field a natural candidate is the photon-number, i.e. Fock basis $\{|m\rangle\langle n|\}$. Under this basis, the process can be expressed as a rank-4 tensor $\varepsilon_{kl}^{mn}$ that relates the density matrix of the output state $\hat{\rho}^{out}$ and that of the input state $\hat{\rho}^{in}$:

$$\rho_{kl}^{out} = \sum_{mn} \varepsilon_{kl}^{mn} \rho_{mn}^{in}, \tag{3}$$

where $\rho_{mn}^{in} = \langle m|\hat{\rho}^{in}|n\rangle$, $\rho_{kl}^{out} = \langle k|\hat{\rho}^{out}|l\rangle$ and $\varepsilon_{kl}^{mn} = \langle k|\varepsilon(|m\rangle\langle n|)|l\rangle$. Characterization of a quantum process can be achieved through the reconstruction of the tensor, which is known as the quantum process tomography (QPT). However, direct reconstruction of $\varepsilon_{kl}^{mn}$ requires superpositions of different Fock states as the probe states, which may be infeasible with current techniques. Luckily, there is another set of complete basis, the coherent states $\{|\alpha\rangle\}$, which is readily generated with the output of a laser. Any quantum state of light $\hat{\rho}$ can be written as a linear combination of density matrices of coherent states $|\alpha\rangle$

$$\hat{\rho}^{in} = 2\int P_{in}(\alpha)|\alpha\rangle\langle\alpha|\, d^2\alpha, \tag{4}$$

where $P_{in}(\alpha)$ is the state's Glauber-Sudarshan $P$ function and the integration is performed over the entire complex plane. The output state can be expressed as:

$$\hat{\rho}^{out} = 2\int P_{in}(\alpha)\, \varepsilon(|\alpha\rangle\langle\alpha|)\, d^2\alpha. \tag{5}$$

Thus, it is sufficient to know the output states of every coherent state $|\alpha\rangle$. In particular the process tensor in Fock basis can be reconstructed using[32]

$$\varepsilon_{kl}^{mn} = 2\int P_{mn}(\alpha)\, \langle k|\,\varepsilon(|\alpha\rangle\langle\alpha|)|l\rangle\, d^2\alpha, \tag{6}$$

where $P_{mn}(\alpha)$ is the $P$ function of $|m\rangle\langle n|$. In practical setups, experimental imperfections and statistical fluctuations are unavoidable. Therefore, instead of directly apply Eq. (6), maximum likelihood estimation (MLE)[36-38] can be used to mitigate of effect of noise.

## 3. Sample and experimental setup

In the experiment, we utilize a typical plasmon-assisted EOT system to investigate the fundamental mechanism inside the photon-plasmon-photon conversion process. The sample is a hexagonal metal-hole array in a 100-nm-thick Au film sputtered on a glass substrate, which is fabricated by using a focused ion beam system (FEI Helios 600i). Each circular hole has a diameter of 460 nm and the array period is 759 nm. The shape of the transmission spectrum of such hexagonal hole array does not depend on the input polarization [39]. Because of the plasmon resonance, the transmissivity of our sample is greatly enhanced to be 62.0% at 1080 nm as shown in Fig. 1(a). The metal holes occupy ~33.4% of the area in our sample. The transmission efficiency $\eta_B$ normalized to the aperture area is 1.86, which indicates that the EOT effect happens [40]. The discrepancy between the experimental data and theoretical simulations is due to the imperfections in the fabrication. We prepare a set of probe states and transmit them through the sample. The quadrature distribution of the output states are measured with a homodyne detector, which allows to reconstruct the Wigner function and density matrix of the output state[41,42] as well as the process tensor.

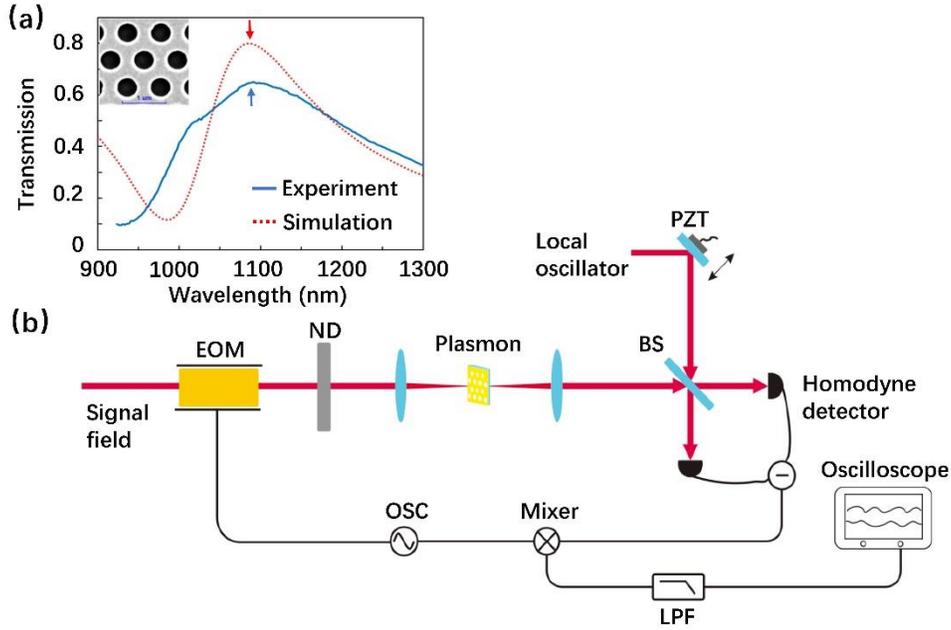

Fig. 1. Transmission spectrum of gold plasmon sample and Experimental setup. (a) The blue slope is EOT transmissivity slope and the red dot is theory transmissivity slope (FDTD simulation). Their characteristic peaks are both at 1080nm. Inset: The electron microscope photo of our metal-hole arrays fragment. Its full size is 65μm×65μm. Period is 759nm. Hole diameter is 460nm. (b) A coherent infrared light at 1080nm is amplitude modulated with an electrooptical modulator and passes through a calibrated neutral density filter to prepare the probe states that incident on the plasmonic sample. The output state is measured with a homodyne detector. The relative phases between the probe states and local oscillator of the homodyne detector are set with a piezoelectric transducer. EOM is amplitude electrooptical modulator. ND is neutral attenuation piece. BS is beam splitter. PZT is piezoelectric transducer. OSC is high frequency signal generator. LPF is low frequency filter.

The schematic experimental setup is shown in Fig. 1(b). The signal field is a coherent infrared light at 1080nm, which is amplitude-modulated by an electro-optical modulator (EOM). The modulation frequency is set at 2 MHz and the applied voltage on the EOM is tunable between 0V and 10V. The modulated signal field is attenuated by a neutral density filter to prepare the probe states $|\alpha\rangle$ with average photon number $|\alpha|^2$ less than 10 and then is focused on the metal hole array sample. A two-lens system is used to reshape the incident beam to match the sample area. For the experiment we use 9 different probe states with the modulation voltage

of the EOM from 0V to 8V with 1V increment, which give $\alpha = 0$, 0.1375, 0.2750, 0.4125, 0.5500, 0.6875, 0.8250, 0.9625, and 1.100, respectively. The transmitted light after the photon-plasmon-photon conversion process is measured with a homodyne detection system. The signal field passes through the metal hole sample and then interferes with a strong local field. The relative phase between the two fields is scanned by a piezoelectric transducer (PZT) from 0 to $2\pi$ and tracked with an ancillary beam. A pair of photodiodes are used to detect the interference intensity and a subtraction is performed on the AC signals. The subtracted AC signal is sent to a lock-in system. After frequency mixing and low-frequency filtering, the signal is collected with an oscilloscope. Such homodyne detection method can efficiently increase the signal-to-noise ratio in our measurement. We can obtain the quadrature information at different angles by slowly scanning the PZT with a frequency of 2Hz.

## 4. Quantum state tomography

First, we remove the sample and perform the quantum state tomography of the input coherent states at various modulation voltages to calibrate the input states and test the system. By scanning the phase of the local field between 0 and $2\pi$, we collect $5\times10^5$ data points, which are equally divided into 20 phase sections. The collected data are used to reconstruct the density matrices $\hat{\rho}^{in}$ in Fock basis of the input states with MLE[43]. The phase-space quasiprobablity distribution, known as the Wigner function can be calculated from $\hat{\rho}^{in}$

$$W(X,Y) = \frac{1}{2\pi}\int_{\infty}^{\infty}\langle X + \frac{1}{2}X'|\hat{\rho}^{in}|X - \frac{1}{2}X'\rangle \exp(-iX'Y)dX', \tag{7}$$

where $X$ and $Y$ are two quadratures.

The Wigner functions of the input states at 2V, 6V, and 8V modulation voltages are shown in Fig. 2(a), (c) and (e), respectively. Clearly, the center of the Wigner function moves away from the origin of the coordinates as increasing the modulation voltage since the amplitude of the input state increases with the modulation voltage. The Wigner functions possess Gaussian shapes and the distances between the centers and the origin are 0.2750, 0.8250, and 1.100, which corresponds to coherent stats $|\alpha\rangle$ with $\alpha = 0.2750$, 0.8250, and 1.100, respectively.

Next, we set up the sample and repeat above procedure to perform the quantum state tomography for the output states after the photon-plasmon-photon conversion. The measured Wigner functions of the output states are shown in Figs. 2(b), (d) and (f), which are corresponding to the input states in Fig. 2(a), (c) and (e), respectively. By carefully examining the input and output Wigner functions, we find the distance from the center of the Wigner function of the output state is reduced compared to that of the input state for all three states, which is due to the attenuation in the conversion process. Moreover, the Wigner function of the output state is rotated by 0.92 radian around the original with respect to the input state, which suggests the conversion process adds a 0.92 radian constant phase to the state. Such differences should be closely related to the mechanism of the plasmon-assisted EOT process.

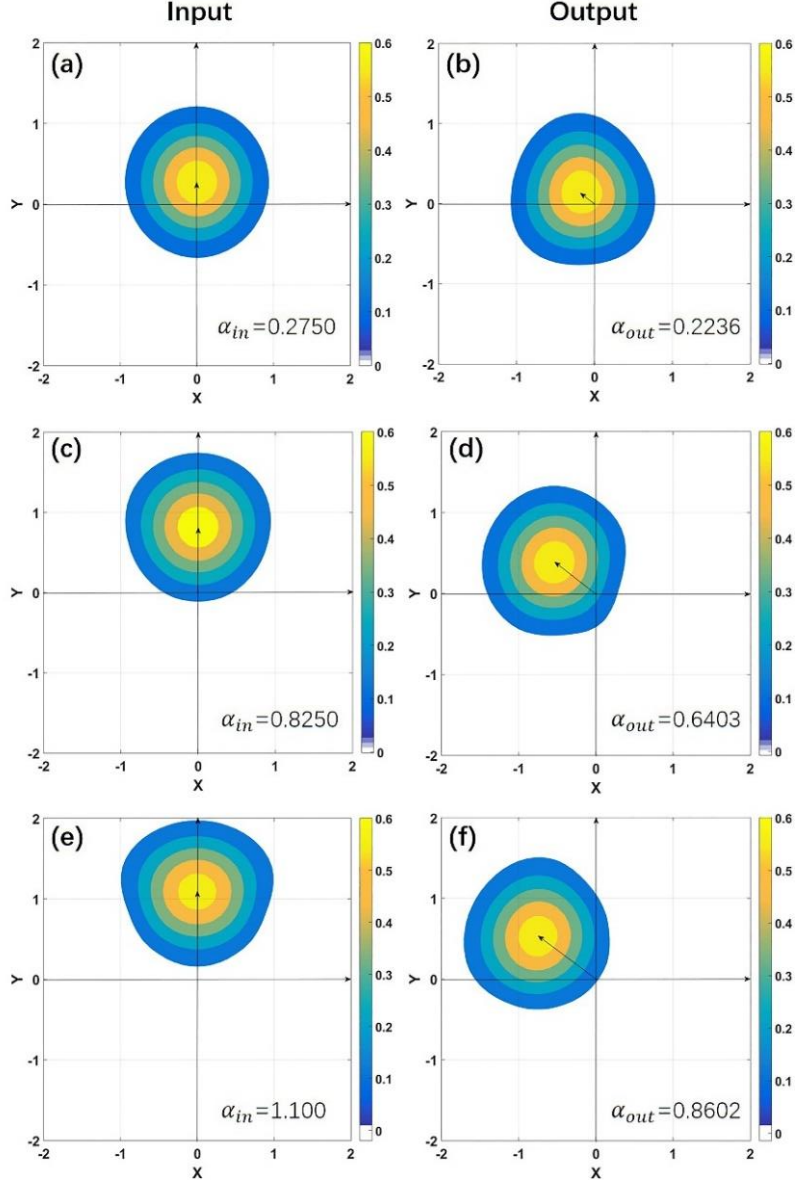

Fig. 2. Wigner function of input and output light. (a/c/e) The Wigner functions of input fields W(X,Y)$_{input}$ with 2V, 6V and 8V modulation voltage of the EOM, respectively. (b/d/f) The Wigner functions of output fields W(X,Y)$_{output}$ with 2V, 6V and 8V modulation voltage of the EOM, respectively. $\alpha_{in}$ corresponds to the input state and $\alpha_{out}$ corresponds to the output state.

## 5. Quantum process tomography of metal-hole arrays

The changes in the Wigner functions of specific input states do not reveal all the information about the conversion process. We further apply CSQPT on the EOT process with 9 coherent states. We collect the outcomes of the homodyne detectors for each output state, and apply MLE algorithm[44] to reconstruct process tensor $\varepsilon_{kl}^{mn}$. Here, $\varepsilon_{kl}^{mn}$ is a four-dimensional tensor. Since the EOT process has influences on both the amplitude and phase characteristics of the input states, we investigate them separately.

We extract the diagonal elements of the tensor $\varepsilon_{kl}^{mn}$ to investigate the amplitude information (Fig. 3a). We choose $m = n$ in the input state and $k = l$ in the output state, which corresponds to the diagonal elements of the density matrices and describes the effect of the process on the photon-number distribution of the state. By analyzing the diagonal elements $\varepsilon_{kk}^{mm}$ of the process tensor, one can see the evolution of photon numbers from input field ($m$) to output field ($k$). For a given input photon number $m$, the output photon number $k$ has a Binomial-like distribution, which hints the Bernoulli transformation of the process. By assuming a linear loss process with transmissivity T = 62.0%, we perform numerical simulations as shown in Fig. 3(b). By comparing the experimental results in Fig. 3(a) and the theoretical simulations in Fig. 3(b), we can get the fidelity F = 99.68% between the two matrices. Clearly, the EOT effect on the amplitude of the input state can be understood as a linear loss.

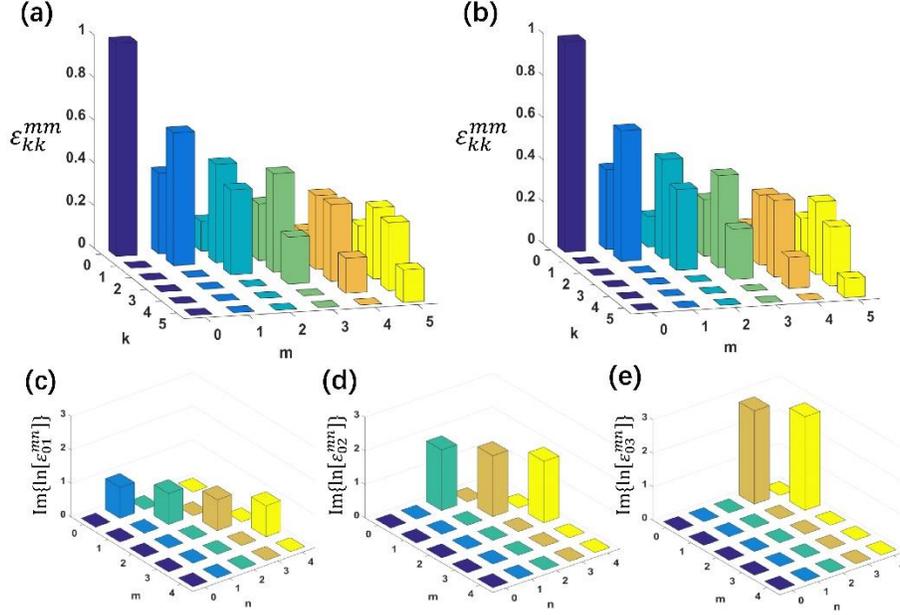

Fig. 3. The results of CSQPT. (a)The diagonal elements of the process tensor $\varepsilon_{kk}^{mm}$ with input field index m and output field index k for an EOT process. (b)Numerical results of a linear loss process with transmissivity of 62.0%. (c/d/e) The off-diagonal elements of the process tensor Im{ln[$\varepsilon_{01}^{mn}$]} (c), Im{ln[$\varepsilon_{02}^{mn}$]} (d), Im{ln[$\varepsilon_{03}^{mn}$]} (e), where m and n represent the input field index in the Fock basis.

In addition to changes in the photon-number distributions, the reconstructed process tensor also reveals the effect on the off-diagonal elements of the density matrix, i.e. the coherence between different photon-number components. In particular, we can acquire the effect of the process on the phases of the input state. Such information can be extracted from the tensor elements mapping the input state density matrix to certain off-diagonal element of the output state. For example, the phase value the $\rho_{01}^{out}$ element of the output state is determined by the phases Im{ln[$\varepsilon_{01}^{mn}$]} of $\varepsilon_{01}^{mn}$ process tensor elements[45]. To elaborate this relation, we decompose the process $\varepsilon$ into a phase shift superoperator and a phase-symmetric process $\varepsilon'$

$$\hat{\rho}^{out} = \varepsilon(\hat{\rho}^{in}) = \widehat{U}(\phi)\varepsilon'(\hat{\rho}^{in})\widehat{U}^\dagger(\phi) = e^{i\phi\hat{a}^\dagger\hat{a}}\varepsilon'(\hat{\rho}^{in})e^{-i\phi\hat{a}^\dagger\hat{a}}, \quad (8)$$

where $\phi$ is a constant phase shift. From Eq. (3), we can get the process tensor $\varepsilon_{kl}^{mn} = |\varepsilon_{kl}'^{mn}|e^{i\varphi_{kl}^{mn}} \cdot e^{i(k-l)\phi}$ and

$$\text{Im}\{\ln[\varepsilon_{kl}^{mn}]\} = \varphi_{kl}^{mn} + (k-l)\phi, \quad (9)$$

where $\varphi_{kl}^{mn}$ is the phase of $\varepsilon_{kl}'^{mn}$, which accounts for photon-number-dependent, i.e. nonlinear, phase shifts.

Fig. 3(c), (d) and (e) show several phase values Im{ln[$\varepsilon_{kl}^{mn}$]} of the EOT process, which are extracted from $\varepsilon_{01}^{mn}$, $\varepsilon_{02}^{mn}$, and $\varepsilon_{03}^{mn}$, respectively. Note for a phase-symmetric process, only the tensor elements $\varepsilon_{kl}^{mn}$ with $k - l = m - n$ are non-zero[32]. The figures show that for given $k$ and $l$, the phase changes are the same for various $m$ and $n$, which well matches the prediction

from the Wigner functions (Fig. 2). These results indicate that there is only a constant linear phase shift involved in the process. In addition, the average values of the phases of $\varepsilon_{01}^{mn}$, $\varepsilon_{02}^{mn}$, and $\varepsilon_{03}^{mn}$ are 0.9200, 1.8182, and 2.7849, respectively, which are linearly dependent on $k-l$. In short, the metal-hole arrays apply a constant phase shift with a value about 0.92 radian to the input state.

## 6. Discussion and conclusion

Although to fully understand the photon-plasmon-photon conversion process requires a detailed microscopic model, the reconstructed process tensor suggests that the conversion process can be effectively described with a beam-splitter model plus a constant phase. With this model, the input field operator $\hat{a}_{in}$ is transferred to the output field operator $\hat{a}_{out}$ through the equation

$$\hat{a}_{out} = \sqrt{1-\eta}\hat{a}_{bath} + e^{i\phi}\sqrt{\eta}\hat{a}_{in}, \tag{10}$$

where $\hat{a}_{bath}$ is the field operator of the lumped Markovian environment in vacuum, $\eta$ is the overall transmissivity and $\phi$ is the constant phase added by the process. The linear-loss model has been proposed in Refs. 8, 19 and tested with squeezed light[19] and heralded single photons[46]. Our tomography results go beyond the previous works by confirming validity of the model for arbitrary input states. Moreover, the new results also highlight the effect of the coupling process on the phase of the input state, which is closely related to quantum coherence. The model given in Eq. (10) can be applied to both classical and quantum optical field, and explains the extraordinary high transmissivity of light and entanglement survival[11,18-20].

To summarize, we have performed complete quantum tomography to reconstruct a photon-plasmon conversion process, i.e., EOT in a metal-hole array. Such reconstruction procedure discovers the fundamental characteristics including both the amplitude and phase information of a typical plasmonic process, which allows us to precisely estimate the interaction of arbitrary classical or quantum optical fields with such plasmonic structure, as well as to develop the microscopic model to interpret the process. Our experimental demonstration provides a fundamental understanding of a plasmon-assisted EOT process, which paves a way for the proper design of a quantum plasmonic components for future applications in on-chip quantum information processing. It should be noted that our experimental observations based on single input mode can be readily extended to the case with multiple optical modes. Since the applications of quantum plasmonics have been further extended to quantum plasmonic sensing and quantum plasmonic networks beyond the classical limit, the complete quantum tomography of various quantum plasmonic process will benefit the optimization of on-chip plasmonic devices.

## Funding

This work was supported by National Key R&D Program of China (2017YFA0303700, 2016YFA0302500), National Natural Science Foundation of China (NSFC) (91636106, 11874213, 11621091, 11474159, 61490711, 11874213).

## References

1. J. A. Schuller, E. S. Barnard, W. Cai, Y. C. Jun, J. S. White, and M. L. Brongersma, "Plasmonics for extreme light concentration and manipulation," Nat. Mater. **9**, 193–204 (2010).
2. E. Ozbay, "Plasmonics: Merging Photonics and Electronics at Nanoscale Dimensions," Science **311**, 189–193 (2006).
3. Y. Zhu, D. Wei, Z. Kuang, Q. Wang, Y. Wang, X. Huang, Y. Zhang, and M. Xiao, "Broadband Variable Meta-Axicons Based on Nano-Aperture Arrays in a Metallic Film," Sci. Rep. **8**, 11591 (2018).
4. Y. Wang, X. Fang, Z. Kuang, H. Wang, D. Wei, Y. Liang, Q. Wang, T. Xu, Y. Zhang, and M. Xiao, "On-chip generation of broadband high-order Laguerre-Gaussian modes in a metasurface," Opt. Lett. **42**, 2463–2466 (2017).

5. D. Wei, Y. Wang, D. Liu, Y. Zhu, W. Zhong, X. Fang, Y. Zhang, and M. Xiao, "Simple and Nondestructive On-Chip Detection of Optical Orbital Angular Momentum through a Single Plasmonic Nanohole," ACS Photonics **4**, 996–1002 (2017).
6. M. S. Tame, K. R. McEnery, Ş. K. Özdemir, J. Lee, S. A. Maier, and M. S. Kim, "Quantum plasmonics," Nat. Phys. **9**, 329–340 (2013).
7. F. Marquier, C. Sauvan, and J.-J. Greffet, "Revisiting Quantum Optics with Surface Plasmons and Plasmonic Resonators," ACS Photonics **4**, 2091–2101 (2017).
8. M. S. Tame, C. Lee, J. Lee, D. Ballester, M. Paternostro, A. V. Zayats, and M. S. Kim, "Single-Photon Excitation of Surface Plasmon Polaritons," Phys. Rev. Lett. **101**, 190504 (2008).
9. Y.-J. Cai, M. Li, X.-F. Ren, C.-L. Zou, X. Xiong, H.-L. Lei, B.-H. Liu, G.-P. Guo, and G.-C. Guo, "High-Visibility On-Chip Quantum Interference of Single Surface Plasmons," Phys. Rev. Applied **2**, 014004 (2014).
10. M.-C. Dheur, E. Devaux, T. W. Ebbesen, A. Baron, J.-C. Rodier, J.-P. Hugonin, P. Lalanne, J.-J. Greffet, G. Messin, and F. Marquier, "Single-plasmon interferences," Sci. Adv. **2**, e1501574 (2016).
11. E. Altewischer, M. P. van Exter, and J. P. Woerdman, "Plasmon-assisted transmission of entangled photons," Nature **418**, 304–306 (2002).
12. X. F. Ren, G. P. Guo, Y. F. Huang, C. F. Li, and G. C. Guo, "Plasmon-assisted transmission of high-dimensional orbital angular-momentum entangled state," EPL **76**, 753 (2006).
13. D. Wang, C. Xia, Q. Wang, Y. Wu, F. Liu, Y. Zhang, and M. Xiao, "Feedback-optimized extraordinary optical transmission of continuous-variable entangled states," Phys. Rev. B **91**, 121406 (2015).
14. A. V. Akimov, A. Mukherjee, C. L. Yu, D. E. Chang, A. S. Zibrov, P. R. Hemmer, H. Park, and M. D. Lukin, "Generation of single optical plasmons in metallic nanowires coupled to quantum dots," Nature **450**, 402–406 (2007).
15. A. Huck, S. Kumar, A. Shakoor, and U. L. Andersen, "Controlled Coupling of a Single Nitrogen-Vacancy Center to a Silver Nanowire," Phys. Rev. Lett. **106**, 096801 (2011).
16. A. I. Fernández-Domínguez, S. I. Bozhevolnyi, and N. A. Mortensen, "Plasmon-Enhanced Generation of Nonclassical Light," ACS Photonics **5**, 3447–3451 (2018).
17. P. Tassin, T. Koschny, M. Kafesaki, and C. M. Soukoulis, "A comparison of graphene, superconductors and metals as conductors for metamaterials and plasmonics," Nat. Photonics **6**, 259–264 (2012).
18. S. Fasel, F. Robin, E. Moreno, D. Erni, N. Gisin, and H. Zbinden, "Energy-Time Entanglement Preservation in Plasmon-Assisted Light Transmission," Phys. Rev. Lett. **94**, 110501 (2005).
19. A. Huck, S. Smolka, P. Lodahl, A. S. Sørensen, A. Boltasseva, J. Janousek, and U. L. Andersen, "Demonstration of Quadrature-Squeezed Surface Plasmons in a Gold Waveguide," Phys. Rev. Lett. **102**, 246802 (2009).
20. B. J. Lawrie, P. G. Evans, and R. C. Pooser, "Extraordinary Optical Transmission of Multimode Quantum Correlations via Localized Surface Plasmons," Phys. Rev. Lett. **110**, 156802 (2013).
21. C. Altuzarra, S. Vezzoli, J. Valente, W. Gao, C. Soci, D. Faccio, and C. Couteau, "Coherent Perfect Absorption in Metamaterials with Entangled Photons," ACS Photonics **4**, 2124–2128 (2017).
22. R. W. Heeres, S. N. Dorenbos, B. Koene, G. S. Solomon, L. P. Kouwenhoven, and V. Zwiller, "On-Chip Single Plasmon Detection," Nano Lett. **10**, 661–664 (2010).
23. A. L. Falk, F. H. L. Koppens, C. L. Yu, K. Kang, N. de Leon Snapp, A. V. Akimov, M.-H. Jo, M. D. Lukin, and H. Park, "Near-field electrical detection of optical plasmons and single-plasmon sources," Nat. Phys. **5**, 475–479 (2009).
24. R. W. Heeres, L. P. Kouwenhoven, and V. Zwiller, "Quantum interference in plasmonic circuits," Nat. Nanotechnol. **8**, 719–722 (2013).
25. J. S. Fakonas, H. Lee, Y. A. Kelaita, and H. A. Atwater, "Two-plasmon quantum interference," Nat. Photonics **8**, 317–320 (2014).
26. S. M. Wang, Q. Q. Cheng, Y. X. Gong, P. Xu, C. Sun, L. Li, T. Li, and S. N. Zhu, "A 14 × 14 μm2 footprint polarization-encoded quantum controlled-NOT gate based on hybrid waveguide," Nat. Commun. **7**, 11490 (2016).
27. C. Lee, F. Dieleman, J. Lee, C. Rockstuhl, S. A. Maier, and M. Tame, "Quantum Plasmonic Sensing: Beyond the Shot-Noise and Diffraction Limit," ACS Photonics **3**, 992–999 (2016).
28. M. W. Holtfrerich, M. Dowran, R. Davidson, B. J. Lawrie, R. C. Pooser, and A. M. Marino, "Toward quantum plasmonic networks," Optica, **3**, 985–988 (2016).
29. S. Fasel, M. Halder, N. Gisin, and H. Zbinden, "Quantum superposition and entanglement of mesoscopic plasmons," New J. Phys. **8**, 13–13 (2006).
30. W. Fan, B. J. Lawrie, and R. C. Pooser, "Quantum plasmonic sensing," Phys. Rev. A **92**, 053812 (2015).
31. M. Dowran, A. Kumar, B. J. Lawrie, R. C. Pooser, and A. M. Marino, "Quantum-enhanced plasmonic sensing," Optica, **5**, 628–633 (2018).
32. M. Lobino, D. Korystov, C. Kupchak, E. Figueroa, B. C. Sanders, and A. I. Lvovsky, "Complete Characterization of Quantum-Optical Processes," Science **322**, 563–566 (2008).
33. T. W. Ebbesen, H. J. Lezec, H. F. Ghaemi, T. Thio, and P. A. Wolff, "Extraordinary optical transmission through sub-wavelength hole arrays," Nature **391**, 667–669 (1998).
34. F. J. García de Abajo, "Colloquium: Light scattering by particle and hole arrays," Rev. Mod. Phys. **79**, 1267–1290 (2007).
35. F. van Beijnum, C. Rétif, C. B. Smiet, H. Liu, P. Lalanne, and M. P. van Exter, "Quasi-cylindrical wave contribution in experiments on extraordinary optical transmission," Nature **492**, 411–414 (2012).

36. A. I. Lvovsky, "Iterative maximum-likelihood reconstruction in quantum homodyne tomography," J. Opt. B: Quantum Semiclass. Opt. **6**, S556–S559 (2004).
37. J. Fiurášek and Z. Hradil, "Maximum-likelihood estimation of quantum processes," Phys. Rev. A **63**, 020101 (2001).
38. M. Ježek, J. Fiurášek, and Z. Hradil, "Quantum inference of states and processes," Phys. Rev. A **68**, 012305 (2003).
39. T. Thio, H. F. Ghaemi, H. J. Lezec, P. A. Wolff, and T. W. Ebbesen, "Surface-plasmon-enhanced transmission through hole arrays in Cr films," J. Opt. Soc. Am. B, JOSAB **16**, 1743–1748 (1999).
40. C. Genet and T. W. Ebbesen, "Light in tiny holes," Nature **445**, 39–46 (2007).
41. G. M. D'Ariano, C. Macchiavello, and M. G. A. Paris, "Detection of the density matrix through optical homodyne tomography without filtered back projection," Phys. Rev. A **50**, 4298–4302 (1994).
42. U. Leonhardt, M. Munroe, T. Kiss, T. Richter, and M. G. Raymer, "Sampling of photon statistics and density matrix using homodyne detection," Opt. Commun. **127**, 144–160 (1996).
43. A. I. Lvovsky and M. G. Raymer, "Continuous-variable optical quantum-state tomography," Rev. Mod. Phys. **81**, 299–332 (2009).
44. S. Rahimi-Keshari, A. Scherer, A. Mann, A. T. Rezakhani, A. I. Lvovsky, and B. C. Sanders, "Quantum process tomography with coherent states," New J. Phys. **13**, 013006 (2011).
45. C. Kupchak, S. Rind, B. Jordaan, and E. Figueroa, "Quantum Process Tomography of an Optically-Controlled Kerr Non-linearity," Sci. Rep. **5**, 16581 (2015).
46. G. Di Martino, Y. Sonnefraud, S. Kéna-Cohen, M. Tame, Ş. K. Özdemir, M. S. Kim, and S. A. Maier, "Quantum Statistics of Surface Plasmon Polaritons in Metallic Stripe Waveguides," Nano Lett. **12**, 2504–2508 (2012).